\documentclass[iop]{emulateapj}
\usepackage{aas_macros}
\usepackage{graphicx}
\usepackage{amssymb}
\usepackage{natbib}
\usepackage[figuresright]{rotating}
\usepackage{color}
\newcommand{\dt}{{\delta\tau}}
\newcommand{\adt}[1]{{\langle\delta\tau_{#1}\rangle}}
\newcommand{\adtbar}[2]{{\langle\overline{\delta\tau_{#1}#2}\rangle}}

\defcitealias{Leka2012}{Paper I}
\defcitealias{Barnes2012}{Paper III}

\begin{document}

\title{Helioseismology of Pre-Emerging Active Regions II: Average Emergence 
Properties}
\author{A.C.~Birch\altaffilmark{1,2}, D.C. Braun\altaffilmark{1}, K.D.~Leka\altaffilmark{1}, G.~Barnes\altaffilmark{1}, B.~Javornik\altaffilmark{1}}
\altaffiltext{1}{Northwest Research Associates, Boulder CO USA}
\altaffiltext{2}{Max-Planck Institut f\"ur Sonnensystemforschung, 37191 Katlenburg-Lindau, Germany}

\begin{abstract}
We report on average sub-surface properties of pre-emerging active
regions as compared to areas where no active region emergence was detected.
Helioseismic holography is applied to samples of the two populations
(pre-emergence and without emergence), each sample having over
100 members, which were selected to minimize systematic bias, as described
in Paper I (Leka et al., 2012). We find that there are statistically
significant signatures (i.e.,\ difference in the means of more than a few standard errors)
 in the average subsurface flows and the apparent wave speed that
precede the formation of an active region.  The measurements here rule
out spatially extended flows of more than about 15~m\,s$^{-1}$ in the top 20~Mm
below the photosphere over the course of the day preceding the start of
visible emergence. These measurements place strong constraints on models
of active region formation.

\end{abstract}

\maketitle

\section{Introduction}\label{sec.introduction}


As discussed in \citet{Leka2012} (Paper I), the mechanisms behind the
formation of solar active regions (AR) are not known.  Possibilities include
magnetic flux tubes rising essentially intact from the base of the convection zone \citep[for a
review see][]{Fan2009}.  Alternatively, near-surface effects could dominate
the formation mechanism \citep[e.g.][]{Brandenburg2005}; hybrid scenarios are
also a possibility.  More generally, understanding of the formation of 
active regions may lead to a better understanding of the solar dynamo. 

Local helioseismology \citep{Gizon2005,Gizon2010} is among the tools
that potentially could be used to
determine the subsurface dynamics associated with active region formation.
While helioseismic analysis samples the state of the plasma below
the visible surface, the interpretation of the results is 
not always straightforward -- especially when the region sampled is quickly 
changing in context, such as an emerging magnetic flux region.  

Numerous studies have attempted to detect -- and thus characterize -- 
the signature of one or two active regions at their earliest stages of formation. 
In one of the earliest studies, \citet{Braun1995} applied Hankel analysis \citep{Braun1987,Braun1992} to 
South Pole observations of the formation of NOAA~AR\,5247.  In the few days 
before the region's sunspot was reported, the Hankel analysis detected 
negative phase shifts (i.e., reduced wave speed).  The cause of the phase 
shifts is not known, and the supporting magnetic measurements were very sparse. 

\citet{Chang1999} used acoustic imaging \citep{Chang1997} applied to MDI and 
also TON data \citep{Chou1995} to study the emergence of NOAA~AR\,7978.  Based on an 
analysis of the focus-depth dependence of observations with a two-day 
resolution, they suggested that they had detected a magnetic flux concentration 
moving upwards towards the photosphere.  However, the detected positive phase shifts that
are co-spatial with the emerging active region are also co-temporal with
the appearance of surface flux.

\citet{Jensen2001} used time-distance helioseismology \citep{Duvall1993} to study two active regions
that formed on 11~January~1998 (NOAA AR\,8131 and AR\,8132) and found changes in
the subsurface wave speed that developed at about the same time as the 
magnetic signatures of the regions were seen in MDI magnetograms.  
There was no apparent change in 
subsurface structure before the surface magnetic features were seen. 

\citet{Hartlep2011} measured the acoustic power from MDI observations before
and during the emergence of AR\,10488.  This study suggested a decrease in acoustic
power in the 3~to~4~mHz band and a reduction in subsurface wavespeed before
significant magnetic flux is seen at the photosphere.

\citet{Kosovichev2009} used time-distance helioseismology
to study two emerging active regions (NOAA AR\,8167 and AR\,10488).  These case studies inferred increases in the subsurface
wave speed associated with flux emergence, however there was no clear
indication that these increases preceded the appearance of magnetic flux
at the photosphere.  \citet{ZharkovThompson2008} also used time-distance helioseismology
in a case-study of the emergence of NOAA AR\,10790;
sound-speed increases were also found with (but not prior to) the 
appearance of surface field.

\citet{Komm2008} used ring-diagram analysis \citep{Hill1988} to infer
the subsurface flows; it was applied to a few active regions undergoing
flux emergence: newly-emerging regions (including AR\,10488) were compared
to older but resurging active regions.  The analysis, which included a
comparison with magnetic field in the target areas, found indications
of upflows prior to emergence that were followed by downflows.

In an extension of the above study to one of the few truly statistical
approaches, \citet{Komm2009,Komm2011} then used the same ring-diagram
analysis approach on a very large sample of active regions characterized
by the evolution of the magnetic field (from increasing to decreasing).
Key to this study was the ``control'' dataset of as many ``quiet'' areas.
This study found that growing active regions show a preference for
upflows of a fraction of m\,s$^{-1}$, with this effect most apparent at the
depth of about 8--10~Mm and reversed below this depth.

There are indications that the details of particular data analysis methods may be very important in determining the helioseismic signals seen for emerging active regions.  \citet{Ilonidis2011} showed several cases (including
AR\,10488) where a travel-time reduction of order 10~s for measurements
sensitive to a depth 60~Mm was seen to precede the emergence of an AR.
\citet{Braun2012} showed that this result is not reproduced (on the same
active regions) using helioseismic holography.   

It is difficult to get a consensus picture of the results, likely
for a variety of reasons.  While most studies report some detection
of a signal associated with the appearance of an active region, only
\citet{Komm2008}, \citet{Hartlep2011}, \citet{Komm2009}, and \citet{Komm2011}
include any comparison to a ``control'' sample by which to evaluate
the detections, although in most cases the criteria for this evaluation are
not discussed in detail.  
The presence of surface fields may complicate the interpretation of 
any `pre-emergence' signal.

While mention of the surface field evolution was included in most studies,
a detailed accounting of the presense (or absence) of surface-field
for the full time period used for the helioseismology data analysis was rarely included.
The studies cited above are predominantly case studies, with the analysis
performed at (or over) varying times in their target regions' evolution.
Case studies are enlightening and critical for guidance -- but as
discussed in \citetalias{Leka2012} and shown in \citet{Ilonidis2011},
regions emerge with different rates and into different surface contexts;
this points to the need for statistical studies, but also reveals why
few strong results have emerged from the large-sample studies to date
\citep{Komm2009,Komm2011}.

Numerical simulations have been used to model the emergence of
magnetic flux through the last few tens of Mm below the photosphere.
\citet{Stein2011} used a radiative magnetoconvection simulation in
which horizontal magnetic field was advected into the simulation domain
by upflows through the lower boundary (20~Mm below the photosphere).
For the case of 10~kG field strength, the magnetic field emerges through
the photosphere in about 32~hours.  In this case, the field has only
a weak effect on the convection and the emergence process is largely
controlled by the convection.  \citet{Cheung2010}, also using a radiative
magnetoconvection simulation, studied the emergence of a semi-torus of
twisted field.  In this case, the magnetic structure was forced through
the lower boundary with a vertical velocity of 1~km~s$^{-1}$ and took
about 2~hours to rise to the photosphere from a depth of 7.5~Mm.  In these
simulations, vertical flows of about 0.5~km~s$^{-1}$ and diverging
horizontal flows of order a few km~s$^{-1}$ are associated with the flux
emergence at the photosphere.  These two simulations show scenarios for
flux emergence with very different observational implications.

In the present study, we combine a statistically-significant sample and
a control group, with a detailed analysis of the surface magnetic field
\citepalias{Leka2012}, and examine the differences in {\it pre-emergence}
behavior in an average sense.  We undertake a more detailed statistical
analysis of the two populations in the next paper in this series
\citep[][Paper III]{Barnes2012}.  It should be noted that the analysis
used herein (see~\S\ref{sec.holography}) relies on ``raw'' travel-time shifts 
rather than inversions used by most of the studies cited above; as such,
we avoid potential complications and uncertainties of the inversions,                
while sacrificing details about the depth of any perturbations detected.
Nonetheless, we show that on average there are
statistically significant sub-surface flows and changes in wave speed that precede
AR emergence; these results place strong constraints on models of active
region formation, even when individual case studies show no clear signature 
of emergence.
 
The layout of the paper is as follows. In \S\ref{sec.data} we review the data 
sets that are used in the helioseismic measurements. In \S\ref{sec.holography} 
we detail the application of helioseismic holography to these data sets.  A 
statistical summary of the results is shown in \S\ref{sec.results}.  We discuss 
the main results and outline some possibilities for future work in 
\S\ref{sec.discussion}.
 
\section{The Data}\label{sec.data}

The GONG Doppler data cubes for active regions before their emergence (we will 
refer to these as pre-emergence, or ``PE'' cases) and for quiet-Sun control 
cases that are not associated with AR emergence (non-emergence, or ``NE'',
cases), accompanied by the associated MDI magnetograms are described in detail by 
\citetalias{Leka2012}. To briefly review the procedure described there:  we 
selected a sample of 107 pre-emergence cases and the same number of 
non-emergence cases. These samples of the two populations (PE and NE) were selected to have 
well-matched distributions in disk location (latitude/CMD) and time (within the 
solar cycle), to avoid biases in the seismology due to projection and instrumental effects.
For the PE cases, a refined emergence time $t_0$ was defined as 
when the change in the total absolute flux (as measured from MDI) reached 10\% 
of the maximum increase detected within a 3-day window of its nominal 
(NOAA-defined) emergence time.   The GONG data cubes were 1664~minutes long 
(one ``GONG-day''), spanning from 1648 minutes before $t_0$ to 16 minutes 
after.  These 1664~minute cubes were divided into five time 
intervals of 384~minutes each, with an overlap of 64~minutes between time 
intervals.  The time between the start of each time interval is 320 minutes.  
The cadence of the GONG Doppler data is one minute per image.

\begin{figure}
\begin{center}
\includegraphics[width=3.5in]{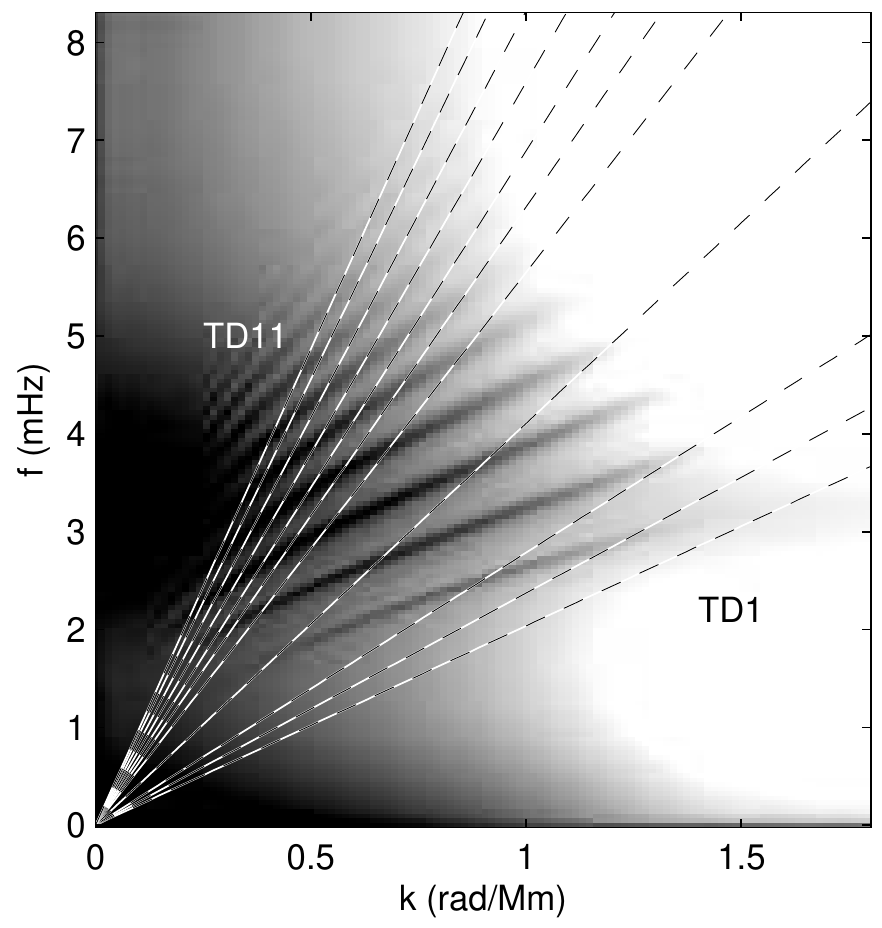}
\end{center}
\caption{Mean power spectrum of the GONG Doppler data cubes for all of the NE 
regions. The ridges are visible up to about horizontal wavenumber 
$k=1.5~{\rm rad\,Mm}^{-1}$.  The central phase speeds for the filters used in this work 
\citep[we use the filters from][]{Couvidat2005} are shown as dashed lines 
(filter TD1 has the lowest phase speed and TD11 has the highest).  The width of 
the filters is similar to the distance between filters. }
\label{fig.mean_power}
\end{figure}

As described by \citetalias{Leka2012}, each of the Doppler images is Postel 
projected \citep{Pearson1990} on to a map with a scale of 1.5184~Mm\,pixel$^{-1}$ and a 
size of 256$\times$256 pixels${}^{2}$. Figure~\ref{fig.mean_power} shows the mean power 
spectrum of the NE regions. The ridges in the power spectrum are visible up to 
about $k=1.5~{\rm rad\,Mm}^{-1}$. 

Potential field extrapolations from MDI magnetograms are used to estimate the 
radial component of the magnetic field \citepalias[for details
see][]{Leka2012}.  These estimates of the radial magnetic field are then 
projected into the same map coordinates as the GONG Dopplergrams and averaged 
over the same time intervals.  The result of this procedure is a set of maps of 
radial magnetic field estimates, one map for each time interval for each PE and 
NE case.
 
 \begin{table}
 \begin{center}
 \begin{tabular}{c c c c}
\hline
filter  & inner radius( Mm) & outer radius (Mm) & depth (Mm) \\           
\hline
TD1     &  3.7 & 8.7 & 1.4 \\
TD2     &  6.2 & 11.2  & 2.2 \\
TD3     &  8.7&  14.5 & 3.2 \\
TD4     & 14.5  & 19.4& 6.2 \\
TD5     & 19.4 & 29.3& 9.5 \\
TD6     & 26.0 &35.1 &  11.4 \\
TD7     & 31.8&  41.7 & 13.3 \\
TD8     & 38.4 & 47.5 &15.7 \\
TD9     & 44.2 & 54.1& 18.2 \\
TD10   & 50.8  & 59.9& 20.9 \\
TD11   & 56.6 &66.7 & 23.3 \\
\end{tabular}
\end{center}
\caption{Table of annulus size and lower turning point depth, taken from Table~1 of \citet{Couvidat2005}.  The first column shows the filter name.
The remaining three columns show the inner annulus radius (Mm), outer  annulus radius (Mm), and the lower turning point depth (computed at the central phase speed of each filter).}
\label{table.filters}
 \end{table}

\section{Helioseismic Holography}\label{sec.holography}

Helioseismic holography \citep{Lindsey2000} is a tool that uses measurements of 
solar oscillations to infer subsurface conditions; it is very similar to 
time-distance helioseismology \citep{Duvall1993}. See \citet{Gizon2005} for a 
detailed review of helioseismic holography.  Here, we applied surface-focusing 
helioseismic holography to each of the five time intervals of each 
of the PE and NE data cubes.

The basic steps in the analysis are: 1)~track and Postel project GONG 
Dopplergrams (described in the previous section), 2)~apply phase-speed filters, 
3)~compute local-control correlations, and 4)~measure travel-time shifts.  
Phase-speed filters \citep{Duvall1997} isolate waves with particular ranges in 
lower turning points.   We used filters 1~through~11 (here denoted as filters 
TD1 to TD11) as described in Table~1 from \citet{Couvidat2005}.  These filters 
cover the range in sampling depth from about 1.4~Mm~(filter TD1) to about 
23.3~Mm~(filter TD11), as shown in Table~\ref{table.filters}.  Each 
phase-speed filter leads to a separate filtered data cube (i.e., filtered time 
series of Dopplergrams).  From each of these filtered data cubes, we then 
computed local-control correlations \citep[e.g.,][these are analogous to 
time-distance correlations]{Lindsey2000} for both center-annulus and 
center-quadrant geometries \citep[see e.g.,][for a description of these 
geometries]{Duvall1997}.   The annulus sizes are shown in Table~\ref{table.filters} and were taken from Table~1 of \citet{Couvidat2005}.  Finally, we measured 
travel-time shifts from the center-annulus and center-quadrant local-control 
correlations using the phase method of \citet{Lindsey2000}.   We emphasize that 
due to the tracking procedure, most of the effect of solar rotation is removed 
from the measurements shown here.  In addition, as described in 
\S\ref{sec.results}, we remove smooth fits to all of the maps of travel-time 
maps.  As a result, the travel-time maps shown here do not include large-scale 
effects (e.g., differential rotation and meridional flow).

Throughout this paper, we use $x$~and~$y$ to denote the coordinates in the 
Postel projection geometry, with $x$ increasing in the direction of rotation 
and $y$ increasing to the north. We use $\dt_{x}$ to denote east-west 
travel-time differences (i.e., the difference in travel times between 
east-going and west-going waves) and $\dt_{y}$ to denote north-south 
travel-time differences.   These travel-time differences are mostly sensitive 
to east-west (north-south) flows, with a negative $\dt_x$ ($\dt_y$) indicative 
of a flow to the west (north).

Following the usual convention, we denote the outgoing (i.e., 
center-to-annulus) travel-time shift (shift relative to the average) by 
$\dt_{\rm out}$ and the in-going (i.e., annulus-to-center) travel-time shift by 
$\dt_{\rm in}$.  From these we construct the ``out minus in'' travel-time shift 
$\dt_{\rm oi}= \dt_{\rm out} -  \dt_{\rm in}$ and the mean travel-time shift 
$\dt_{\rm mn}= \left[\dt_{\rm out} +  \dt_{\rm in}\right]/2$.  The one-way 
center-annulus travel times $\dt_{\rm in}$ and $\dt_{\rm out}$ are sensitive to 
the isotropic wave speed and also to vertical flows and converging/diverging 
horizontal flows.  These two effects (i.e., change in isotropic waves speed and 
the presence of flows) are approximately separated by  $\dt_{\rm oi}$, which is 
sensitive mostly to  diverging / converging horizontal flows and vertical flow, 
and $\dt_{\rm mn}$ which is more sensitive to changes in the isotropic wave 
speed.  The sign conventions are such that $\dt_{\rm mn}<0$ is interpreted as a 
signature of increased wave speed, and $\dt_{\rm oi} <0$ is the signature of a 
diverging flow.

From the quantities $\dt_x$ and $\dt_y$ we computed proxies for the vertical 
component of the flow vorticity (denoted vort) ${\rm vor} = \partial_x \dt_y - \partial_y \dt_x$ 
and for the horizontal flow divergence (denoted div) as ${\rm div} = \partial_x \dt_x + \partial_y \dt_y$. 
The derivatives were computed in the Fourier (horizontal wavevector) domain.

\section{Results}\label{sec.results}

The analysis is presented in three stages: first, individual
examples; second, averages over all samples (separately for
the two populations) but retaining the spatial information;
and third, an analysis of the spatial average over a small
central region, and its temporal evolution, again contrasting
the averages over the two populations.

\subsection{Individual Active Region Examples}\label{sec.egs}

\begin{figure*}
\centerline{\includegraphics[width=\linewidth]{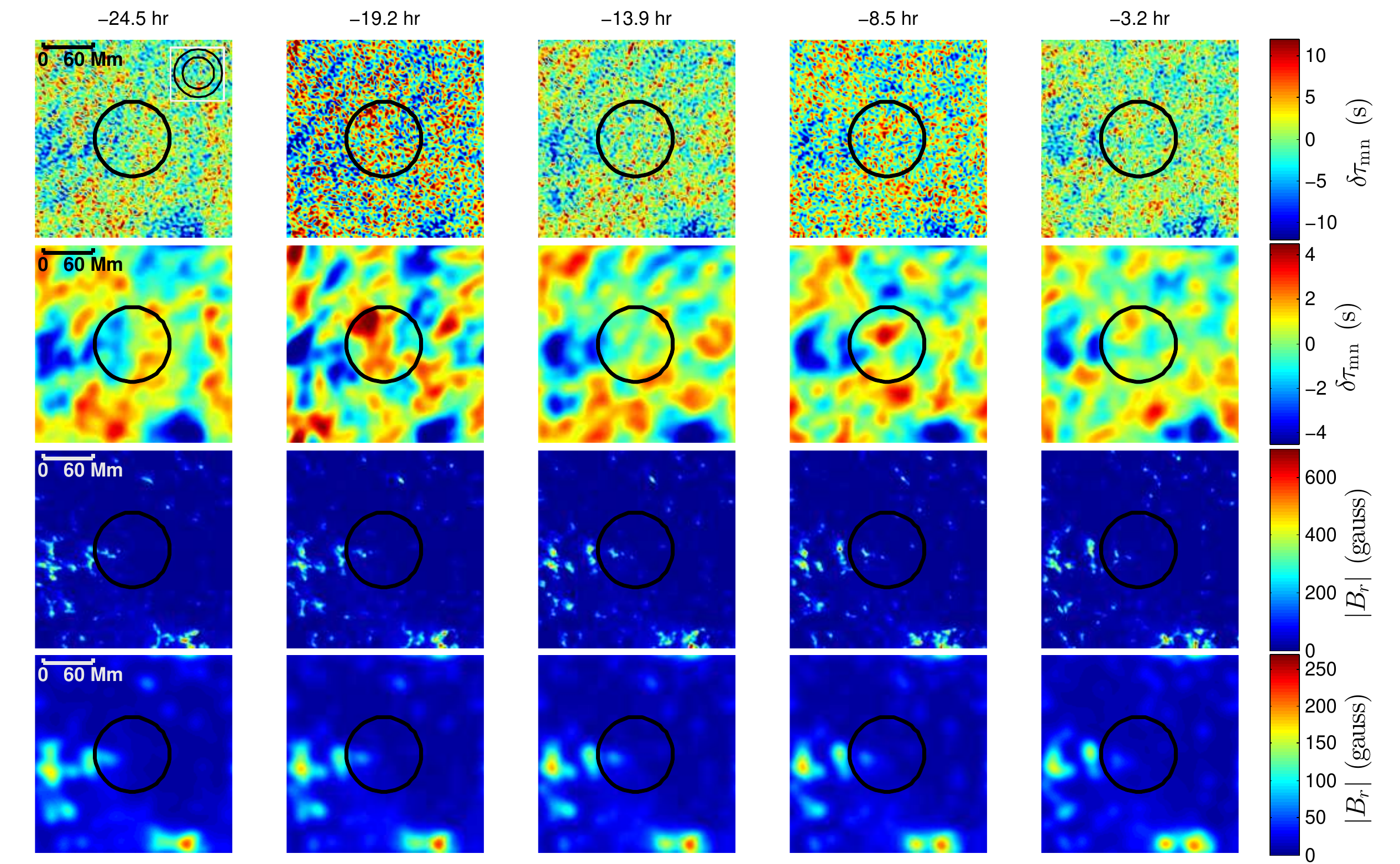}}
\caption{Example, for NOAA AR\,9729, of the time evolution of the
$\dt_{\rm mn}$ and of the surface magnetic field over the 24~hr prior
to the appearance of the region's emerging magnetic flux.  Time advances 
from left to right in steps of 320 minutes; the mid-interval time
 relative to $t_0$ is labeled at the top.  {\bf Top row:} maps of 
$\dt_{\rm mn}$ for filter TD5 (corresponding to a depth of about 9.5~Mm) 
for each of the time intervals.  The last column covers
$-368{\rm min} \leq t_0 \leq 16{\rm min}$, little if any active-region
flux emergence should be visible.  {\bf Second row:} same as the top
row after smoothing with a Gaussian with full-width at half maximum
(FWHM) of 9.99 pixels (about 15~Mm).  Notice the change in color scale;
the spatial smoothing reduces the amplitude of the fluctuations in
$\dt_{\rm mn}$.  {\bf Third row:} the evolution of the unsigned radial
magnetic field strength.  {\bf Fourth row:} same, but after the same
smoothing as was applied to the maps of $\dt_{\rm mn}$; again, the color scale 
changes because of the reduced amplitude of the fluctuations.  Notice that
negative travel-time shifts appear co-spatial with the areas of strongest
magnetic field.  The black circle in all panels shows the averaging
region (see \S~\ref{sec.spatialaverages}) surrounding the area where the
AR will emerge.  The inset (top left panel) indicates the size of the
annulus used in the holography measurements for filter TD5.  Notice the
higher noise level in $\dt_{\rm mn}$ maps for the second time interval; this
is caused by a low (60\%) duty cycle.  As discussed in \S\ref{sec.sampleaverage}, such low duty cycle time intervals 
are excluded in later analysis.}
\label{fig.example_AR9729}
\end{figure*}

Figure~\ref{fig.example_AR9729} shows, for the case of the emergence of 
AR\,9729, the $\dt_{\rm mn}$ maps for filter TD5 (lower turning point of about 
9.5~Mm) together with estimates of the radial magnetic field obtained from 
potential field extrapolations (details are in \citetalias{Leka2012}) from MDI 
96-minute magnetograms.   In this case, the emergence takes place in a region 
with strong nearby magnetic fields.  These pre-existing surface magnetic fields have 
corresponding negative features (i.e.\ increased apparent wave speed) in the 
maps of $\dt_{\rm mn}$.  Using the NE cases, we estimate that the noise level in the smoothed maps (for maps with duty cycle of better than 70\%) of $\dt_{\rm mn}$ for filter TD5 is about 0.9~s.
Therefore, many of the weaker features seen in the travel-time shifts in Figure~\ref{fig.example_AR9729} are likely due to noise. With this noise level in mind, there is no apparent evolution with time.  Notice that this figure covers roughly the 24 hours before emergence, and as a result the development of the active region is not seen in either the magnetograms or the travel-time maps. 
This is a typical example of the PE cases.

\begin{figure*}
\centerline{\includegraphics[width=\linewidth]{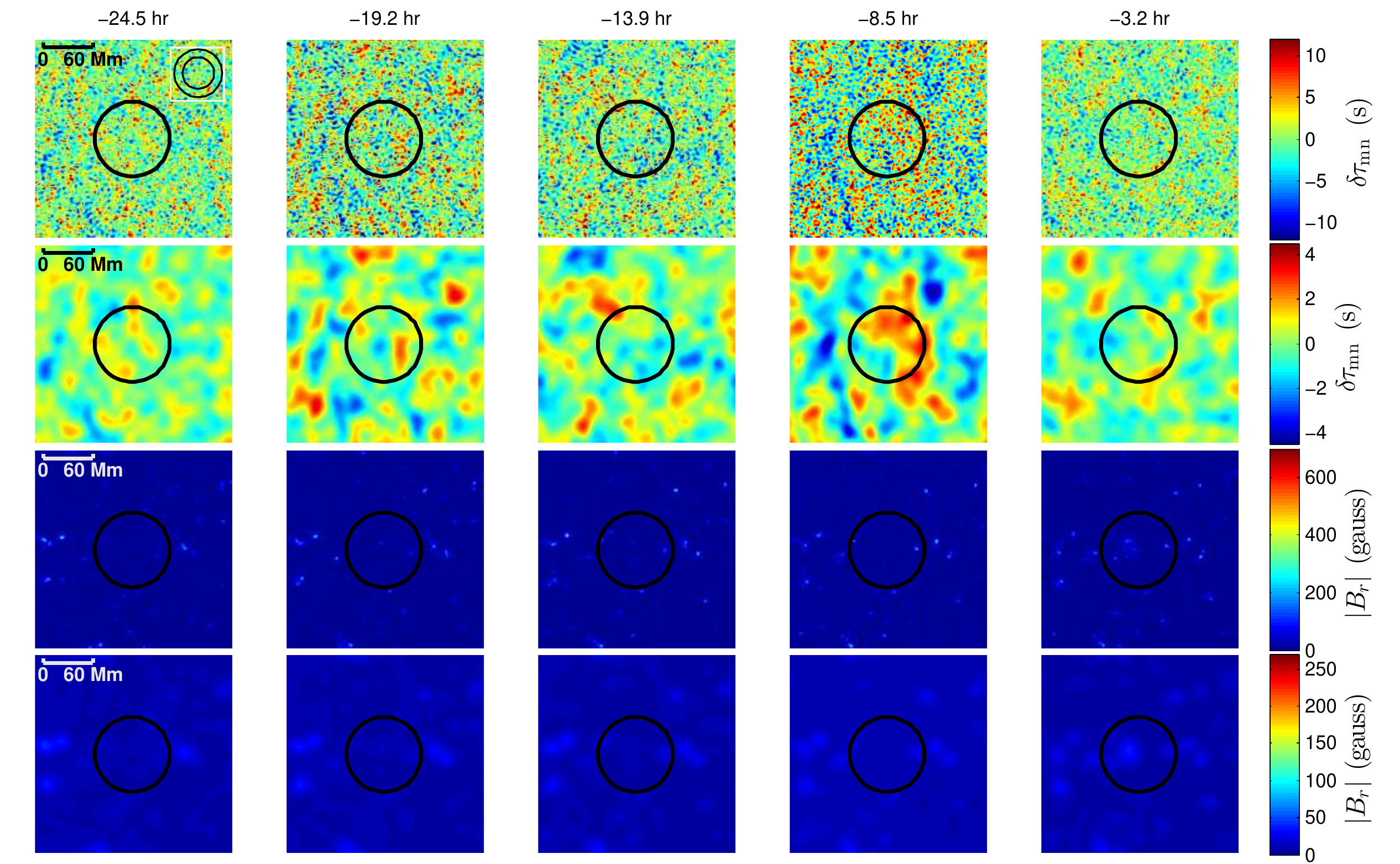}}
\caption{Same as Figure~\ref{fig.example_AR9729}, but for AR\,10488. As in
Figure~\ref{fig.example_AR9729}, notice that the color scale is different for
the first and seconds rows, and for the third and fourth rows.  In this case,
there is no clear signature of emergence in the magnetic field or $\dt_{\rm
mn}$ maps in the 24~hours before emergence.}
\label{fig.example_AR10488}
\end{figure*}

Figure~\ref{fig.example_AR10488} shows another example, this time for the
pre-emergence time evolution of NOAA~AR\,10488, that was also studied by
\citet{Komm2008}, \citet{KosovichevDuvall2008}, \citet{Kosovichev2009},
\citet{Hartlep2011}, and \citet{Ilonidis2011}.  In our analysis, there is no
discernable signal greater than $\pm 2\ {\rm s}$ in the first four time intervals
(covering approximately $-27\ {\rm hr} \leq t_0 \leq -5\ {\rm hr}$) in the raw or
smoothed $\dt_{\rm mn}$ maps; the raw and smoothed magnetic flux maps are
similarly bare.  In fact, there are fewer areas of nearby strong field and
spatially-associated $\dt_{\rm mn}$ signal than in the previous example
(Figure~\ref{fig.example_AR9729}).

In this case, a feature with $\approx50$~G magnitude appears in the average
magnetogram for the last time interval.  This is not strictly unusual; the
automated definition of emergence time that we have used, while refined earlier
by hours or days from the NOAA-assigned emergence time, may still allow
magnetic field at the surface prior to the ``emergence time'' $t_0$ of
11:11\,UT on 26 October 2003 in this case \citepalias[see][for
details]{Leka2012}.  Here we note, however, that less objective definitions of
emergence lead to less repeatable results; flux emergence does not always
follow a standard template for temporal evolution.   Nonetheless, the signal is
weak in this last time interval (it is an average over $\approx6$~hr) and has
no corresponding features in the maps of $\dt_{\rm mn}$, though this may be due
mostly to noise (which we expect, based on the statistics of the NE cases, to
be about 0.9~s in the smoothed travel-time maps for the filter shown in
Fig.~\ref{fig.example_AR10488}).

Of the other studies of this region, all use a different emergence time; here,
we summarize the results relative to our definition of emergence time.
\citet{Komm2008} begin their analysis on 27 October 2003, and thus there is no
possibility for a direct comparison of the results, as this is the day
following our analysis.  \citet{KosovichevDuvall2008,Kosovichev2009} first see
wave speed perturbations in data from an 8~hr time interval centered at
approximately one hour after $t_0$, and describe this as a pre-emergence
signature with growth of the magnetic flux starting at approximately
$t_0+9$~hr.  Thus, the results are similar to this study in that no clear
subsurface signal appears when only data from at least a few hours prior to
$t_0$ are used in the analysis.

\citet{Hartlep2011} state that magnetic field starts to appear at the surface
approximately 2~hr before $t_0$, while significant magnetic flux appears at
around $t_0$.  Their finding of a change in the acoustic power in the 3--4\,mHz
frequency range, averaged over 128 minutes, centered approximately two hours
before $t_0$ is likely related to the weak magnetic field seen in the last time
interval of this study.

Finally, \citet{Ilonidis2011} report a reduction in the mean travel-time of
about 16~s at a depth of about 60~Mm (much deeper than the 9.5~Mm shown here)
from an 8~hr data set centered approximately 7.5~hr before $t_0$.  The
difference in the depths considered again makes a direct comparison impossible,
but see \citet{Braun2012} for a discussion of using helioseismic holography to
examine the same depth range.

Even for this most frequently studied active region emergence, little direct
comparison is possible due to different time intervals and depth ranges
covered.  However, in the most comparable studies of
\citet{KosovichevDuvall2008} and \citet{Kosovichev2009}, the results are
similar to the results presented here when the difference in the stated time
of emergence is accounted for.

\subsection{Averages for the Two Populations}\label{sec.sampleaverage}

We now present the results of averages over all members of a 
sample (e.g., the average over all NE cases) with duty cycle greater than 70\%
(c.f., \citetalias{Leka2012}, Table 3; all averages in this paper will use this constraint on the duty cycle).  Averages taken
in this manner are indicated by angle 
brackets, $\langle\cdot\rangle$. 
For the travel-time maps, large-scale spatial
variations resulting from the Postel projection have been 
removed prior to averaging, using a second order polynomial fit in the $x$ and $y$ map coordinates 
(the functional form of the fitting polynomial is $f(x,y)=a_1 x^2 + a_2 y^2 + 
a_3 x y + a_4 x + a_5 y +a_6$).  For this analysis, all maps were smoothed with a Gaussian of FWHM of 
9.99 pixels. In the case of the PE samples, the emergence site was not initially at the center of every data cube.
To improve the signal/noise ratio for this analysis, we 
aligned all of the travel-time maps (and magnetograms) before averaging.   As 
described by \citetalias{Leka2012}, the alignment was done by first computing, 
for each PE case, the centroid of the pixels where the change in time averaged 
$B_r$ between time interval~0 (about 27 to 21 hours before emergence) and the average 
of $B_r$ computed over the 4 to 10 hours after emergence was more than 30\% of 
the maximum.  

\begin{figure*}
\centerline{\includegraphics[width=\linewidth]{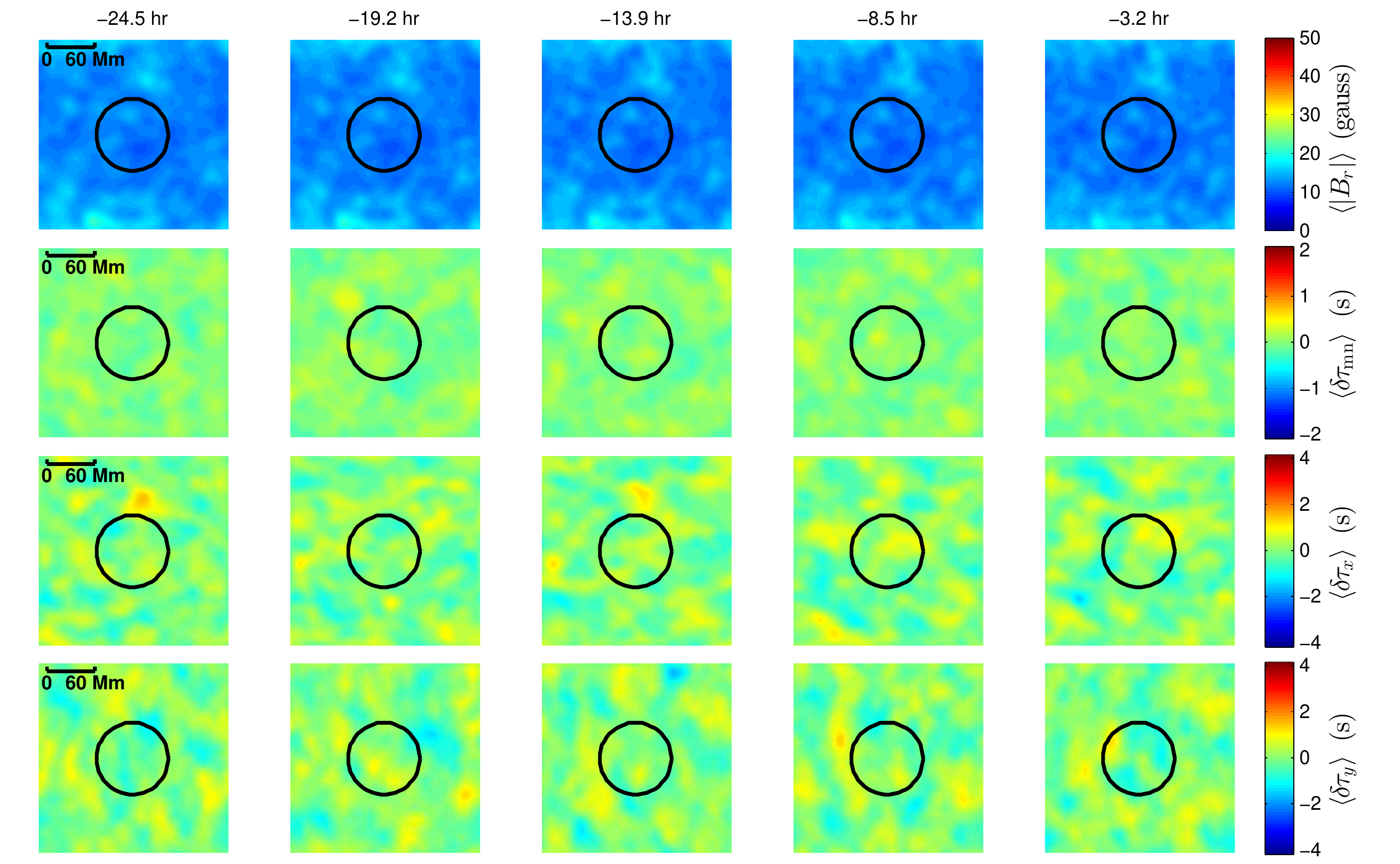}}
\caption{Evolution (left to right, times labeled as Figure~\ref{fig.example_AR9729})
of $\langle|B_r|\rangle$, $\adt{\rm mn}$,$\adt{x}$, and $\adt{y}$ maps for filter TD5,
of the average for each time interval taken over all NE samples with duty cycle great than 70\%.
No clear and persistent features are visible, and no features are present
that are distinctly greater than the noise. The black circle in each panel 
indicates the area that is spatially averaged in subsequent analysis.}
\label{fig.average_maps_qs}
\end{figure*}

Figure~\ref{fig.average_maps_qs} shows the time-evolution
of $\adt{\rm mn}$, $\adt{x}$, and
$\adt{y}$ maps over all of the NE cases, for filter TD5.
Also shown is the evolution of the corresponding average of the unsigned
radial magnetic field.  The average magnetic field is overall weak and
unchanging over these averages.  There are regions where the average
magnetic field is above 20~Gauss, generally near the edges of the map.
Nearby plage field was permitted in the definition of a ``non-emerging''
region \citepalias[see][for a discussion]{Leka2012}.  The maps of $\adt{\rm
mn}$, $\adt{x}$, and $\adt{y}$, all for filter TD5, show no features
stronger than $\pm 1{\rm s}$, and little spatial or temporal coherence
between the time intervals.  

\begin{figure*}
\centerline{\includegraphics[width=\linewidth]{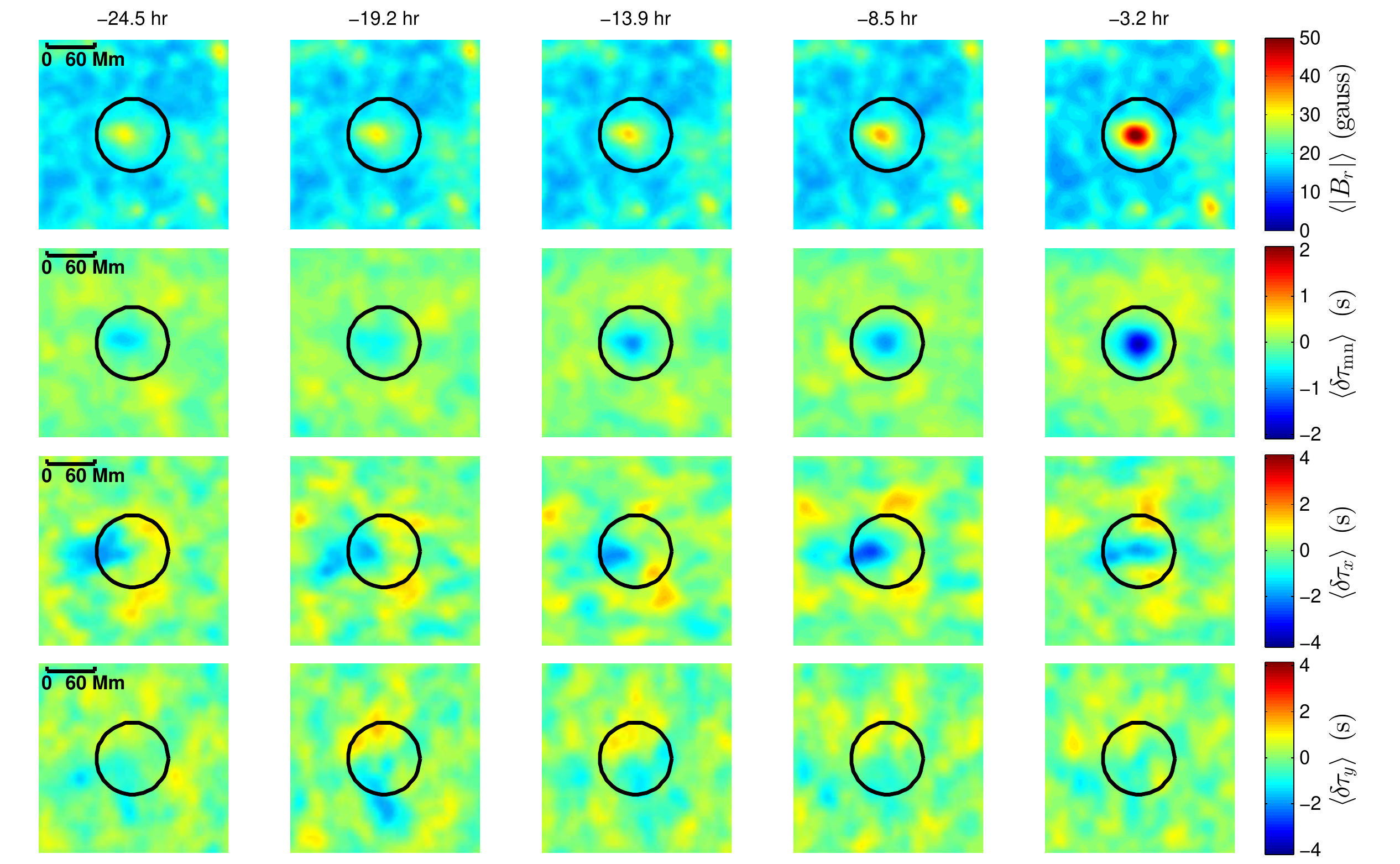}}
\caption{Same as Figure~\ref{fig.average_maps_qs}, showing ({\bf top: bottom})
$\langle|B_r|\rangle$, $\langle\dt_{\rm mn}\rangle$, $\langle\dt_x\rangle$, and 
$\langle\dt_y\rangle$ but for the 
average over all of the PE samples (with duty cycle great than 70\%) again for filter TD5. 
The features in the maps of $\adt{\rm mn}$ are spatially 
correlated with those in the maps of $\langle|B_r|\rangle$. For the first four 
time intervals, there are anti-symmetric features in the travel-time differences $\adt{x}$ and 
$\adt{y}$; these features might be hints of a converging flow.  
The map of $\adt{x}$ for the last time intervals shows what might be evidence of a 
weak prograde flow associated with the surface magnetic field 
\citep[e.g.][]{Zhao2004}.  These travel-time differences are of order two 
seconds, which for filter TD5 corresponds to a flow of about 15~m\, s$^{-1}$.}
\label{fig.average_maps_ar}
\end{figure*}

Figure~\ref{fig.average_maps_ar} shows the time-evolution of the same 
averaged quantities for the PE samples
(spatially co-aligned as described above).  The difference between
these results and those shown in Figure~\ref{fig.average_maps_qs}
is striking.
There is a persistent feature of order -1~s in the maps of $\adt{\rm mn}$ for 
filter TD5, with the strength of the signal 
increasing as $t_0$ approaches. This signal appears to be 
associated with a feature in the average magnetogram in the same location. In 
general, we see correspondence between strong features in the average radial 
magnetic field strength and features in the mean travel-time shifts.  One 
possibility is that the surface magnetic field is the direct cause of the 
travel-time shift \citep[i.e., the showerglass effect of][]{Lindsey2005}.  We 
undertake a statistical study of this possibility in the next paper in this 
series \citep[][Paper III]{Barnes2012}.  

We also see (Fig.~\ref{fig.average_maps_ar}) that the average radial field 
strength at the emergence site also increases as $t_0$ approaches.
While the definition of $t_0$ allows nearby flux to be present 
prior to $t_0$, and may be uncertain to a few hours (see \citetalias{Leka2012}),
a signal can be seen at least a day before emergence.
There is a clear preference for surface field to be located at the emergence site 
prior to significant flux emergence (see \citetalias{Leka2012} and \citetalias{Barnes2012}
for further discussion).

Figure~\ref{fig.average_maps_ar} also shows there are persistent features
in the maps of $\adt{x}$ and $\adt{y}$ with amplitudes of about 2~s
beginning a day prior to $t_0$.  For the first four time intervals, there are hints of anti-symmetric features in these maps; 
a flow converging towards the emergence site at approximately
15~m\,s$^{-1}$ might give such a distribution of travel times.  The maps for the last time
interval are more complicated; in particular the map of $\adt{x}$ shows
what might possibly be the signal of the magnetic region moving in the prograde
direction \citep[e.g.,][]{Zhao2004}.

\subsection{Spatial Averages for the Two Populations}\label{sec.spatialaverages}

Motivated by Figure~\ref{fig.average_maps_ar}, and to focus more
closely on the site of flux emergence, we next present spatial
averages.  The averages are computed, for each sample in the
PE and NE populations, over disks of radius 45.5~Mm centered on
the emergence location (as defined using the centroid mentioned
above, c.f. details in \citetalias{Leka2012}), and indicated
in Figures~\ref{fig.example_AR9729}--\ref{fig.average_maps_ar}.
We denote these initial spatial averages with an over-bar, e.g.,
$\overline{\delta\tau_x}$.   As discussed earlier, data cubes with duty cycle less than 70\%
are omitted from this analysis (c.f., \citetalias{Leka2012}, Table 3).
We then compute averages over the entire PE and NE samples, separately
for each time interval, and the standard error in the mean ($\sigma/\sqrt{n}$, 
where $\sigma$ is the standard deviation and $n$ is the number of samples included). 
The results,
described further below, are depicted in a multi-panel presentation
that includes the parameters, all filters, and the temporal evolution
of both PE and NE averages and their errors within each filter/parameter
combination.

\begin{sidewaysfigure*}
\vspace*{2.8in}
\centerline{\includegraphics[width=\linewidth]{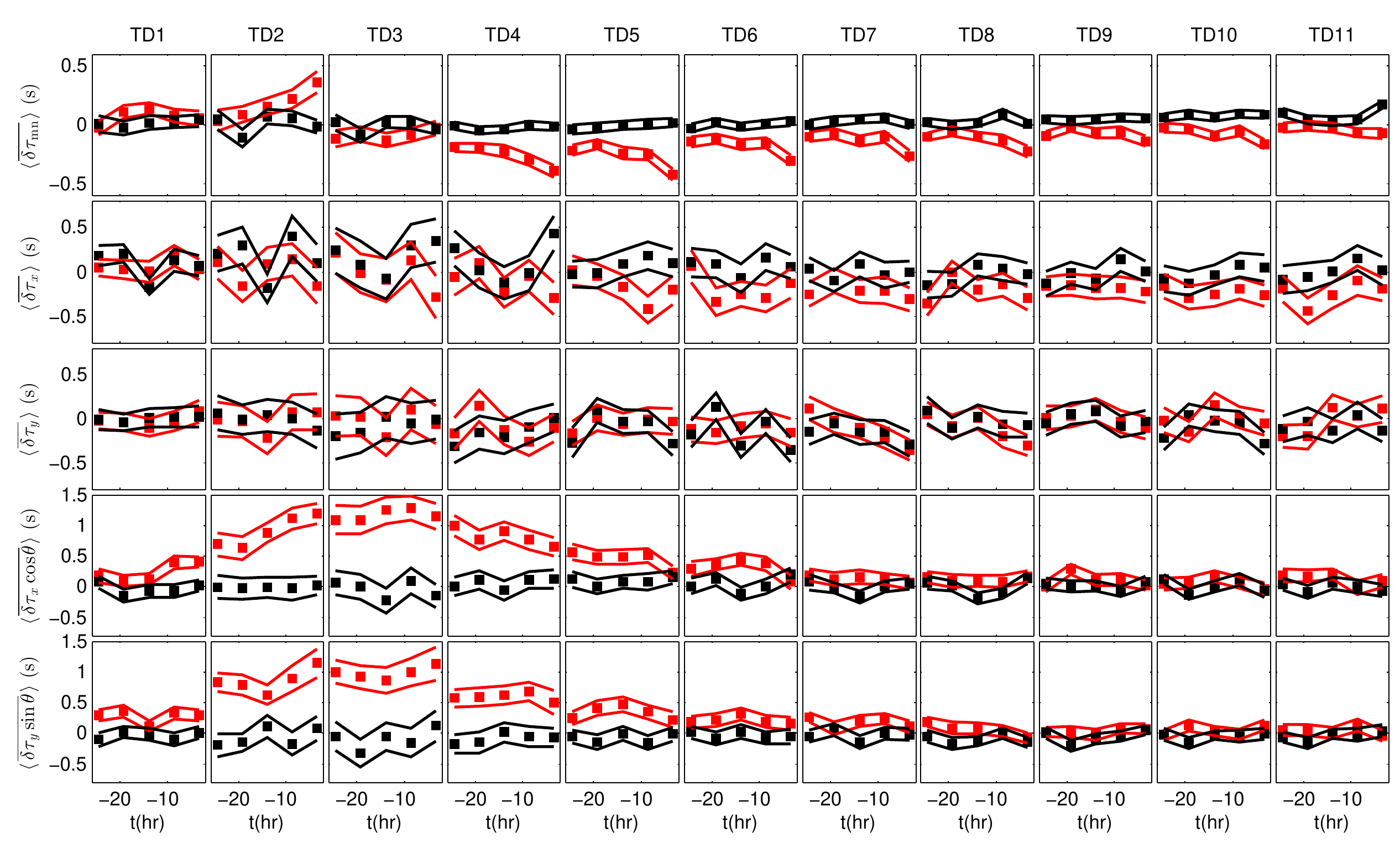}}
\caption{Summary presentation of ({\bf top:bottom}) $\adtbar{\rm mn}{}$, $\adtbar{x}{}$, 
$\adtbar{y}{}$,  $\adtbar{x}{\cos{\theta}}$,  and $\adtbar{y}{\sin{\theta}}$ 
(see text for description); shown are the average results for both PE (red) and NE (black) 
spatial averages, as a function of time (x-axis). 
Each column ({\bf left:right}) corresponds to a different phase-speed filter, with depth of the lower 
turning point increasing from left to right (see Table~\ref{table.filters}).  
The range surrounding each point indicates the associated errors.
These results are discussed in the text.}
\label{fig.confusogram}
\end{sidewaysfigure*}

Figure~\ref{fig.confusogram} shows the time evolution of $\adtbar{\rm
mn}{}$, $\adtbar{x}{}$, and $\adtbar{y}{}$.  The figure also shows
$\adtbar{x}{\cos{\theta}}$ and $\adtbar{y}{\sin{\theta}}$; these
quantities are the sample average (angle brackets) of the spatial
averages (over-bar) of either $\dt_x$ weighted by $\cos{\theta}$
or $\dt_y$ by $\sin{\theta}$, where $\theta$ is the angle measured
counter-clockwise from the direction of solar rotation (the $+\hat{\bf x}$
direction).

Focusing first on the measurements  $\adtbar{\rm mn}{}$, there appear
to be clear and sustained differences between the PE and NE cases for
all but the shallowest filters (TD1, TD2, and TD3), and the deepest filter 
(TD11).  The sense of the difference is generally that $\adtbar{\rm mn}{}$
is more negative for the PE cases than the NE cases.  The NE cases
are very temporally consistent between filters, and in most filters the
difference between the NE and PE samples increases as time approaches
the emergence time $t_0$, with the PE population results increasing in
(negative) magnitude.  The amplitude of this effect is of order tenths
of a second.  This may perhaps be a consequence of the surface magnetic
field and is discussed in detail by \citetalias{Barnes2012}.

Regarding the flow measurements $\adtbar{x}{}$ and $\adtbar{y}{}$,
there is no clear difference between the NE and PE populations.
The anti-symmetric averages $\adtbar{x}{\cos{\theta}}$ and
$\adtbar{y}{\sin{\theta}}$, however, show substantial differences
(e.g.\ of between two and three standard errors for filter TD3) for
filters TD2-TD5 (depths of 2.2~Mm to 9.5~Mm).  These differences
are due to the anti-symmetric features seen in Figure~\ref{fig.average_maps_ar}.  Notice that in all of these
filters (TD2-TD5) these features are seen even at 24~hours before emergence. 

\begin{sidewaysfigure*}
\vspace*{2.8in}
\centerline{\includegraphics[width=\linewidth]{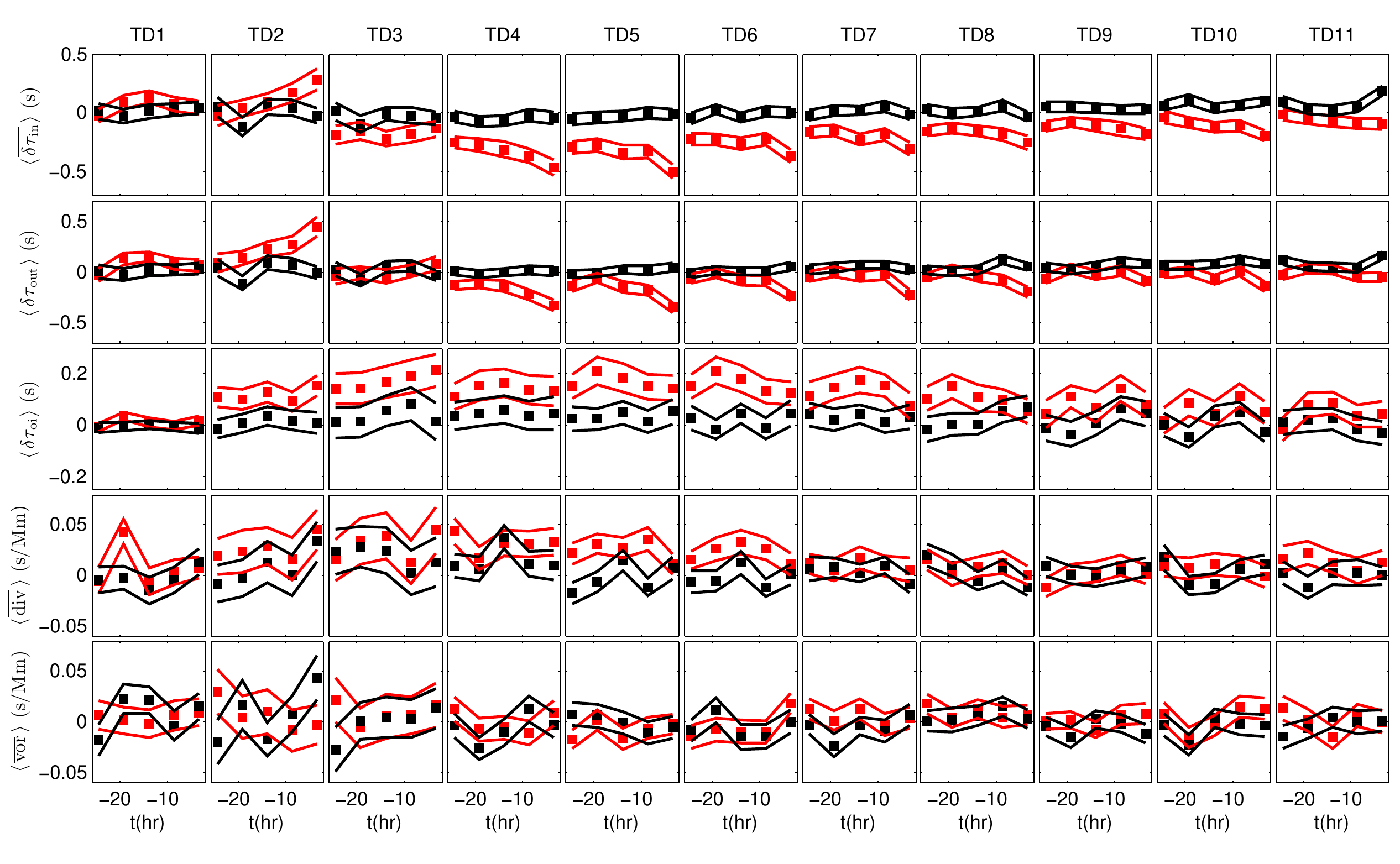}}
\caption{Temporal evolution of $\adtbar{\rm in}{}$, $\adtbar{\rm out}{}$, 
$\adtbar{\rm oi}{}$, $\langle\overline{\rm div}\rangle$, and 
$\langle\overline{\rm vor}\rangle$ for the PE (red) and NE (black) populations. 
 The layout of the figure is the same as for Figure~\ref{fig.confusogram}.}
\label{fig.confusogram2}
\end{sidewaysfigure*}

Figure~\ref{fig.confusogram2} shows the time evolution of $\adtbar{\rm
in}{}$, $\adtbar{\rm out}{}$, $\adtbar{\rm oi}{}$, $\langle\overline{\rm
div}\rangle$, and $\langle\overline{\rm vor}\rangle$.  The parameters
$\adtbar{\rm in}{}$ and $\adtbar{\rm out}{}$ show fairly consistent
differences between the PE and NE samples.  These appear to
be roughly consistent with the change in $\adtbar{\rm mn}{}$ seen
in Figure~\ref{fig.confusogram}.  Similarly, there is a trend that
$\adtbar{\rm oi}{}$ is systematically larger in the PE than the NE cases,
again this is qualitatively consistent with the maps of $\adt{x}$ and $\adt{y}$ from Figure~\ref{fig.average_maps_ar}.

Other results are less pronounced.  For example, in $\langle\overline{\rm
div}\rangle$ there is a weak difference between the NE and PE samples
for filters TD4, TD5, and TD6; the sign is consistant with a converging
flow.   Other PE/NE differences are either fleeting or at the one-$\sigma$
level.  For example, the average divergence $\langle\overline{\rm
div}\rangle$ shows a strong positive feature for filter TD1 in the second
time interval, this feature is not persistent in time and may be noise.
In the first time interval, there are differences, at the one-$\sigma$
level, between the  $\langle\overline{\rm vor}\rangle$ for the PE and NE
cases for filters TD2, TD3, TD5, and TD11.  There are also one-$\sigma$
differences for the last time interval for filter TD2, the second time
interval for TD7, and the last time interval for TD9.

In the interest of recognizing spurious results, we note the following.
Assuming that the variances of two distributions are equal (e.g., PE
and NE; the assumption is reasonable if the variance is predominantly
due to noise), simply by chance alone approximately 16\% of the results should have the
difference between the means greater than the sum of the standard
errors (i.e., $|\bar{x}_1 - \bar{x}_2| > \sigma_1/\sqrt{n_1} +
\sigma_2/\sqrt{n_2}$, where $\sigma$ is the standard deviation, and
as mentioned above, $\sigma/\sqrt{n}$  is the standard error shown in
Figures~\ref{fig.confusogram} and~\ref{fig.confusogram2}). 
This translates to roughly nine $1-\sigma$ results per parameter
that are expected to be purely spurious.  For the difference being twice the sum
of the standard errors, the expectation shrinks to 0.5\%, or less than
one result per parameter, and for the difference being three times the
sum, the expectation would be a nearly negligible $3\times10^{-3}$\%.
For the case of $\langle\overline{\rm vor}\rangle$, there are seven such 
points.  Hence, it seems possible that the differences we have seen in  
$\langle\overline{\rm vor}\rangle$ and $\langle\overline{\rm div}\rangle$
as well as some of the other differences between the PE/NE samples are 
simply the result of noise.  Conversely, the many large differences in 
$\adtbar{\rm in}{}$, for example, are very likely real.

\subsection{Results for the ``Ultra-Clean'' Pre-Emergence Sample}\label{sec.clean}

In \citetalias{Leka2012} we describe a subset of the PE sample which we 
deemed ``ultra-clean''.  These 11 hand-selected emerging active regions
all eventually reached a size of at least 70~$\mu\!\!$~H and displayed
a monotonically increasing flux history after a distinct and unambiguous 
emergence time $t_0$, appearing into an otherwise weak field area.

\begin{figure*}
\centerline{\includegraphics[width=\linewidth]{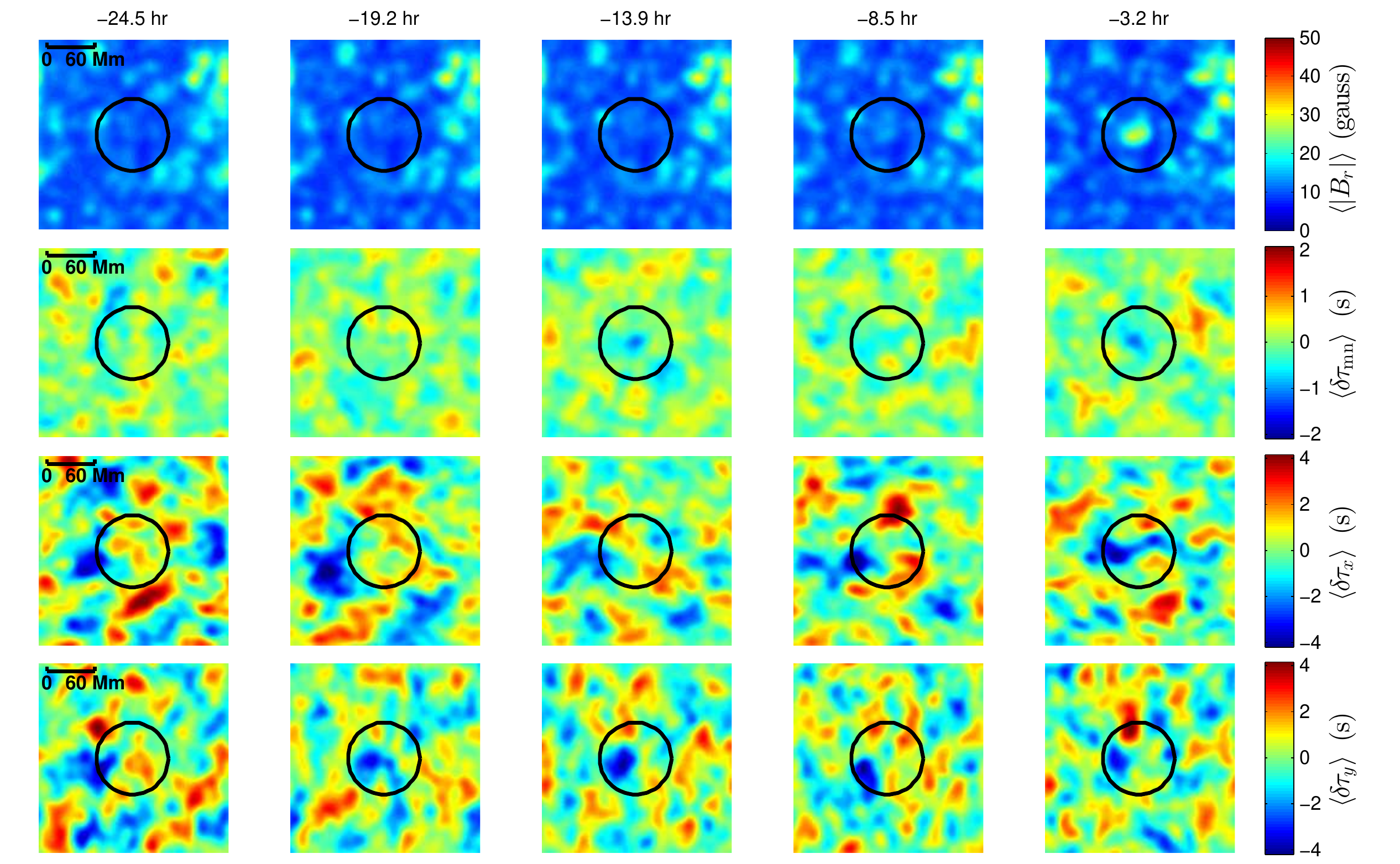}}
\caption{Same as Figure~\ref{fig.average_maps_qs}, but for an average over the 
eleven hand-selected ``ultra-clean'' PE regions.  In this case, the 
$\langle|B_r|\rangle < 20~{\rm G}$ at the emergence site until the last time 
interval.   The noise level is higher in the $\adt{\rm mn}$, $\adt{x}$, and 
$\adt{y}$ compared with the averages shown in Figure~\ref{fig.average_maps_ar}, 
as the average is over a much smaller sample of regions (roughly nine instead of ninety, see Table~3 of Paper I).  There is a 
very weak signal in $\adt{\rm mn}$ at the time intervals centered at -13.9~hr
and -3.2~hr.}
\label{fig.subset_b_average}
\end{figure*}

Figure~\ref{fig.subset_b_average} shows $\langle|B_r|\rangle$,
$\adt{\rm mn}$, $\adt{x}$, and $\adt{y}$ averaged over this subsample
of the PE cases.  As in the average over all PE cases, the strongest
magnetic feature appears at the emergence site in the last time interval.
Also as in the average of all PE cases, there is an apparently associated
feature in the average of $\dt_{\rm mn}$.  
Interestingly, a similar feature in the
-13.9~hr interval has no corresponding feature in $\langle|B_r|\rangle$,
and there is no $\adt{\rm mn}$ feature in the interim -8.5~hr interval.
Due to the smaller number of regions, the noise is higher in these
averages (we expect the noise to be about $0.9/\sqrt{10} \approx 0.3$~s in the smoothed maps of $\adt{\rm mn}$), thus, it is not clear if lack
of features in the average $\dt_{\rm mn}$ map at the earlier times is
significant.  The $\adt{x}$ and $\adt{y}$ maps are also noisier than the
corresponding maps in Figure~\ref{fig.average_maps_ar}, as expected.
There is still a weak suggestion in some of the time intervals of
anti-symmetric features in the average maps of $\dt_x$ and $\dt_y$
(again as in Fig.~\ref{fig.average_maps_ar}).  
Still, it is worth  emphasizing that these 11 hand-picked regions presumably represent the best-case (i.e.\ relatively simple emergence into quiet Sun) scenario for detecting helioseismic signals in the day prior to an active region's emergence.

\begin{sidewaysfigure*}
\vspace*{2.8in}
\centerline{\includegraphics[width=\linewidth]{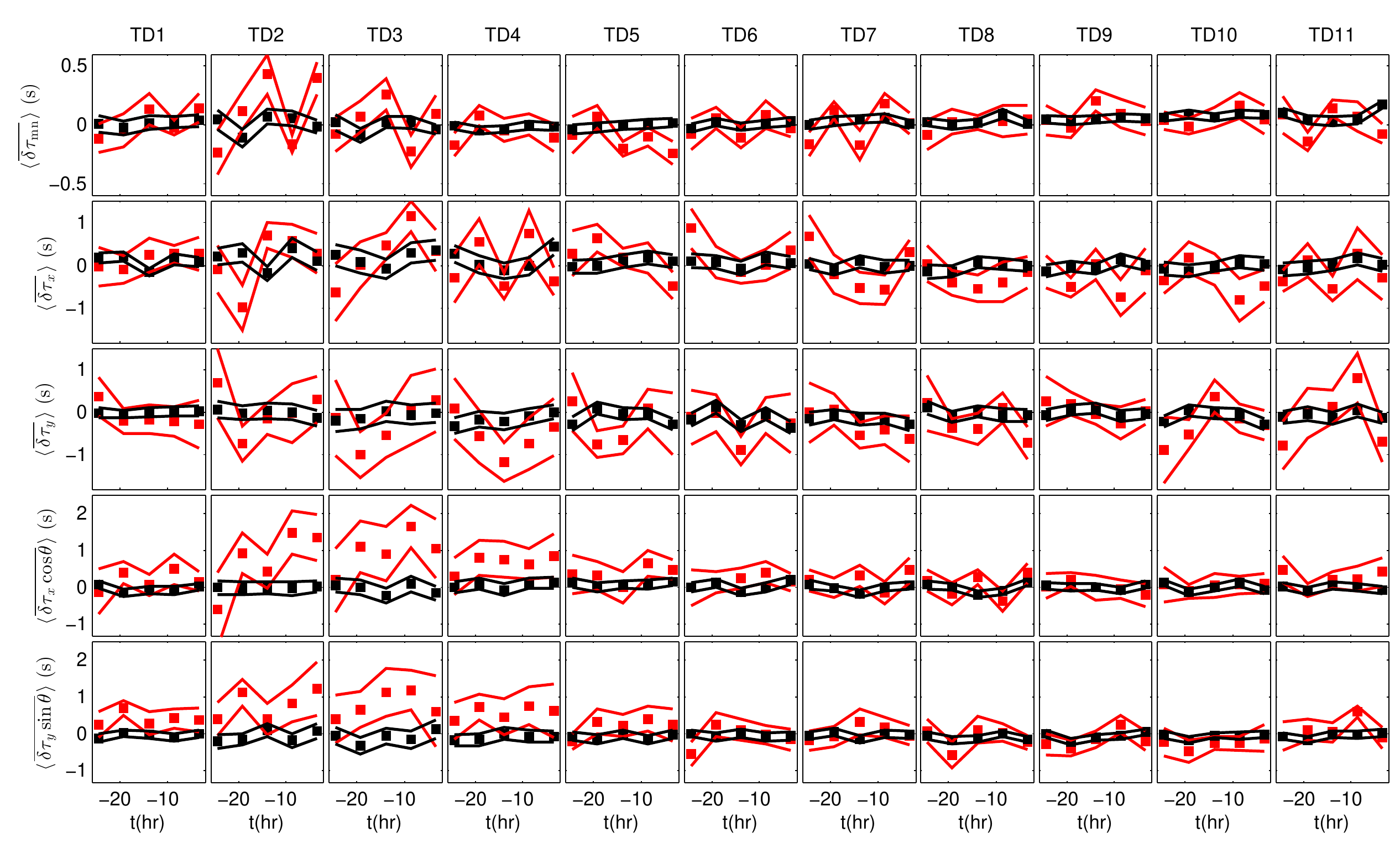}}
\caption{Spatial and ensemble averages $\adtbar{\rm mn}{}$, $\adtbar{x}{}$, 
$\adtbar{y}{}$, $\adtbar{x}{\cos{\theta}}$, and $\adtbar{y}{\sin{\theta}}$  for 
the subsample of eleven PE regions (red) and all of the NE regions (black).   
Notice that the plotting ranges for all variables except $\adtbar{\rm mn}{}$ 
are different than in Figure~\ref{fig.confusogram}.}
\label{fig.confusogram_b}
\end{sidewaysfigure*}

Figure~\ref{fig.confusogram_b} shows the spatial and ensemble averages 
$\adtbar{\rm mn}{}$, $\adtbar{x}{}$, $\adtbar{y}{}$, 
$\adtbar{x}{\cos{\theta}}$, $\adtbar{y}{\sin{\theta}}$  for the subsample of 
eleven PE regions, along with the same average over the NE regions shown in 
Figure~\ref{fig.confusogram}.  The results are less clear than what is presented 
in Figure~\ref{fig.confusogram}, due to the much smaller sample size
for the PE population (about nine instead of 90, after accounting for time-intervals with low duty cycle, see Table~3 of Paper~I).  For this subsample of pre-emergence 
active regions, no difference 
between the $\adtbar{\rm mn}{}$ is seen.  This may be because in the 
hand-selected subsample of PE cases, there is very little nearby surface magnetic 
field except in the last time interval (see Fig.~\ref{fig.subset_b_average}). 
There is no apparent signal in  $\adtbar{x}{}$ or $\adtbar{y}{}$.  The 
differences between the PE and NE populations can, however, still be seen in the 
anti-symmetric averages  $\adtbar{x}{\cos{\theta}}$, $\adtbar{y}{\sin{\theta}}$ 
for filters TD2-TD5.  It is noteworthy that these differences remain, even when 
the differences in  $\adtbar{\rm mn}{}$ are not visible.  This 
suggests that there are different physical mechanisms responsible for these two 
effects.

\begin{sidewaysfigure*}
\vspace*{2.8in}
\centerline{\includegraphics[width=\linewidth]{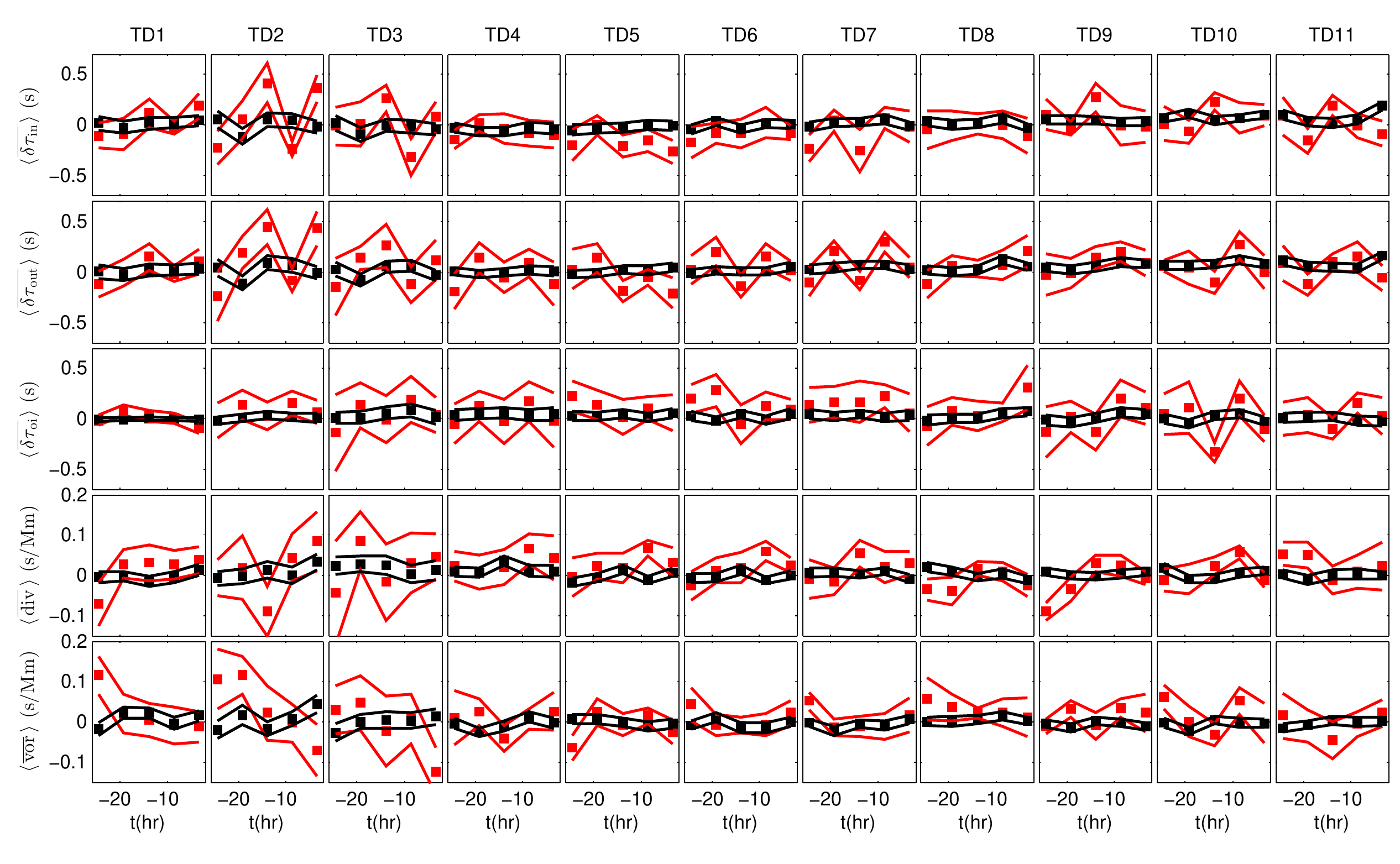}}
\caption{Spatial and ensemble averages  $\adtbar{\rm in}{}$, $\adtbar{\rm 
out}{}$, $\adtbar{\rm oi}{}$, $\langle\overline{\rm div}\rangle$, and 
$\langle\overline{\rm vor}\rangle$ for the subsample of eleven PE regions (red) 
and all of the NE regions (black).  Notice that the plotting ranges for all 
variables except  $\adtbar{\rm in}{}$ and $\adtbar{\rm out}{}$ are different 
than in Figure~\ref{fig.confusogram2}.}
\label{fig.confusogram2_b}
\end{sidewaysfigure*}

Figure~\ref{fig.confusogram2_b} shows the spatial and ensemble averages  
$\adtbar{\rm in}{}$, $\adtbar{\rm out}{}$, $\adtbar{\rm oi}{}$, 
$\langle\overline{\rm div}\rangle$, and $\langle\overline{\rm vor}\rangle$ for 
the subset of eleven PE regions, following
Figure~\ref{fig.confusogram2}.  For the variables  $\adtbar{\rm in}{}$, 
$\adtbar{\rm out}{}$, $\adtbar{\rm oi}{}$, and $\langle\overline{\rm 
div}\rangle$ there are no statistically significant differences in 
the means for the two populations.  For the case of  $\langle\overline{\rm 
vor}\rangle$, there is still a trend to see differences at up to the 
two-$\sigma$ level for the first time interval, with the sign of the effect 
depending on the filter.  

For these results, we note again that the standard errors
for the subset of the PE sample (Figures~\ref{fig.confusogram_b} and~\ref{fig.confusogram2_b})
are much larger than those shown for the full PE 
sample in Fig.~\ref{fig.confusogram2}.  In addition, due to the smaller sample
size, the chances of spurious differences at and beyond the $1-\sigma$ level
are significantly larger  
at approximately 21\% of the results. 
This translates to roughly twelve $1-\sigma$ results per parameter
that are expected to be purely spurious.

\section{Discussion}\label{sec.discussion}

We have applied helioseismic holography to 107 pre-emergence active regions and an 
equal number of areas where no active region emergence occurred.  The 
sample of emerging active regions, as described in \citetalias{Leka2012}, spans 
a wide range of eventual sizes and temporal evolution profiles of 
$B_r$; in addition, these active regions emerge into a variety of magnetic contexts.   
We present single-region examples, but concentrate herein primarily on 
averages over the samples in the two populations, and on spatial
(and ensemble) averages over the central emergence area.

We have found differences in the average seismic signatures between these
two populations.  The pre-emergence regions show a $\dt_{\rm mn}$ that is
reduced by several tenths of a second (in all but the shallowest layers),
compared to the non-emergence population.  There are persistent anti-symmetric features
with magnitudes of up to 2~s in $\dt_x$ and $\dt_y$ in the shallow filters.  These features may qualitatively suggest a converging flow of order 15~m\,s$^{-1}$.   No clear statistically significant differences between the average pre-emergence sample and no-emergence sample were found in the vorticity and only hints seen in the  divergence.

However, two caveats must be included in the above summary.  First,
there exists a weak but persistent average magnetic signal
at the emergence site prior to detectable flux emergence; it strengthens
as the emergence start approaches, as does the magnitude of the co-spatial
$\dt_{\rm mn}$ deficit.  The causal relationship between these features is 
not clear:  does the change in wave speed and the apparent flow result from the 
surface magnetic field? has the magnetic field collected at the emergence site 
because of the converging flows?   

Another possibility is that this is another example of a bias in the selection
of the samples used in this investigation, and our results are a consequence of
emergence happening preferentially at the boundary between supergranules.  Our
non-emergence sample has no preferential location compared to supergranules,
but if the emergence locations are preferentially centered on the boundary
between supergranules, this would result in flows towards the emergence
location.  The flows would not then be converging on the emergence location,
but rather the emergence location would be between neighboring diverging flows
from the supergranules.  

The depth of the antisymmetric features in $\adt{x}$ and $\adt{y}$ approximately
corresponds to the typical depths to which supergranular flow is detected, 
$\approx 5$~Mm, and a
typical supergranule lifetime of 1-2~days is at least as long as the time
considered here \citep{RieutordRincon2010}.  Further, the presence of surface
magnetic field prior to emergence could also be a result of the tendency for
magnetic flux to concentrate in the boundaries between supergranules. Finally,
this might also account for the comparatively weak signature of emergence in
$\adtbar{\rm oi}{}$, when compared to the signature in
$\adtbar{x}{\cos{\theta}}$, and $\adtbar{y}{\sin{\theta}}$, as this is not a
simple converging flow.  Thus we may not be seeing a signal of emergence so
much as a signal of supergranules.  This in itself would be an interesting
result, as it demonstrates a preference for emergence to occur at the boundary
between supergranules. 

We note, however, the typical flow speed for a supergranule of several hundred
meters per second \citep{RieutordRincon2010} is an order of magnitude larger
than our ensemble average of observed travel time shifts.  This could be a
result of imperfect alignment between the boundary and the emergence location,
keeping in mind that our averaging disk is large compared to a typical
supergranule size. By averaging over multiple supergranules for each emergence,
the average travel time difference would be greatly reduced.  The difference in
flow speeds could also result from weakening flow toward the edge of the
supergranule.  

In \citetalias{Barnes2012} we attempt to disentangle the relationships among
some of the parameters where a difference between PE and NE samples is found.
For example, do the $\dt_{\rm mn}$ maps contain information that is not in the
maps of $|B_r|$?  In addition, \citetalias{Barnes2012} uses Discriminant
Analysis \citep{Kendall1983} to determine which measurements are best able to
distinguish the PE and NE populations, as a complement to the ensemble averaging
performed here.

Second, when a subset of eleven ``best'' active region candidates
(emerging with a fast, monotonically-increasing flux history into
an essentially field-free area) were considered, some of the differences described above
disappeared.  Specifically, there were no longer detectable differences
in $\adtbar{\rm mn}{}$ or $\adtbar{\rm oi}{}$.  However, the direction-weighted 
$\adtbar{x}{\cos \theta}$ and $\adtbar{y}{\sin \theta}$ parameters continue to show a statistically significant
difference (of up to 2~s), for moderate depths.  This subset includes
the much-studied NOAA~AR\,10488 (see \citetalias{Leka2012} for the full list).

The results presented here place strong constraints on models of the emergence 
of AR.  We have shown that in the $\approx 24$~hours before emergence 
that the sub-surface flows, averaged  over 6~hours, are no more than $\approx15~{\rm m\,s}^{-1}$.  This raises the question of the time evolution of the strong (100~m\,s$^{-1}$) retrograde flows suggested by models 
\citep[e.g.][]{Fan2008}: why have we not observed these flows? how do these 
retrograde flows interact with the near-surface shear layer?

It is not clear how to reconcile the results presented here with the strong 
pre-emergence reduction (of order 10~s) in the mean travel-time 
seen by \citet{Ilonidis2011}.  \citet{Ilonidis2011} suggest that 
this signature may be caused by a rising flux concentration crossing upwards 
through the depth of 60~Mm, where their analysis is sensitive.  Here, we use measurements that are sensitive to depths shallower than about 20~Mm, and see no signature in $\dt_{\rm mn}$ larger 
than a fraction of a second.  Further measurements are needed to connect these 
two conclusions; why does the helioseismic impact of the rising tube all but disappear 
as the tube approaches the photosphere?

We have found that, on average for our pre-emergence sample, there are
surface magnetic fields present at the emergence site but which vary
only slowly \citepalias[see Fig.~10 of][]{Leka2012} over the day before
flux emergence begins in earnest.   As discussed above, this may be due to a preference for emergence to occur at the boundaries between supergranules where ``quiet'' magnetic flux typically accumulates continuously.
As another possibility, we speculate that this may perhaps
be a result of the interaction between convection and the rising magnetic
fields, i.e., that the portion of the flux that is caught in the fastest
upflows arrives at the surface well before the bulk of the flux tube.
Yet another possibility may be that flux emergence occurs preferentially
into remnant field. A more detailed analysis is needed to disentangle
these effects.

The data analysis we have presented here suggests that the rapid emergence 
process simulated by \citet{Cheung2010} is not typical.  Horizontal flows of 
order km~s$^{-1}$ extending over tens of Mm would produce signals in $\dt_x$ 
and $\dt_y$ of order tens of seconds --  well above our noise level.  
Notice, however, if flows of this strength were to develop only {\it after} the 
emergence time, they would not appear in the current analysis.  The same is true 
for the retrograde flows seen in the rising flux tube simulations discussed in
\citet{Fan2008}.  The picture we 
find here is apparently more compatible with the scenario of \citet{Stein2011}, 
in which weak field is brought to the surface by convection, which is itself 
only weakly altered by the magnetic field.

\acknowledgments 
The authors acknowledge support from NASA contract NNH07CD25C. 
This work utilizes data obtained by the Global Oscillation Network Group 
(GONG) Program.  GONG is managed by the National Solar Observatory, which is 
operated by AURA, Inc.\ under a cooperative agreement with the National Science 
Foundation.  The data were acquired by instruments operated  by the Big Bear 
Solar Observatory, High Altitude Observatory, Learmonth Solar Observatory, 
Udaipur Solar Observatory, Instituto de Astrofísica de Canarias, and Cerro 
Tololo Interamerican Observatory.  MDI data were provided by the SOHO/MDI consortium;
SOHO is a project of international cooperation between ESA and NASA.  A.C.B. acknowledges collaborative work within the framework of the DFG CRC 963 ``Astrophysical Flow Instabilities and Turbulence."

\bibliographystyle{apj}

\end{document}